\documentclass[preprint,5p, anonymous=true, sort&compress]{elsarticle}
\usepackage[utf8]{inputenc}
\usepackage{graphicx, amsmath, textcomp}
\usepackage{url, hyperref}
\hypersetup{colorlinks=true}
\usepackage{subfigure}
\usepackage{placeins}
\usepackage{booktabs}
\usepackage{multirow}
\usepackage{eurosym}
\usepackage{algorithm}
\usepackage{algpseudocode}
\usepackage{amssymb, bm, bbold}

\usepackage{enumitem}
\setlist{leftmargin=20pt} 

\usepackage{array}
\usepackage[table]{xcolor}
\newcolumntype{L}[1]{>{\raggedright\let\newline\\\arraybackslash\hspace{0pt}}m{#1}}
\newcolumntype{C}[1]{>{\centering\let\newline\\\arraybackslash\hspace{0pt}}m{#1}}
\newcolumntype{R}[1]{>{\raggedleft\let\newline\\\arraybackslash\hspace{0pt}}m{#1}}

\newcolumntype{H}{>{\setbox0=\hbox\bgroup}c<{\egroup}@{}}

\usepackage[nameinlink]{cleveref}
\crefname{figure}{Fig.\hspace{-1pt}}{Fig.\hspace{-1pt}}
\Crefname{figure}{Figure}{Figures}
\crefname{equation}{Eq.\hspace{-1pt}}{Eq.\hspace{-1pt}}
\Crefname{equation}{Equation}{Equations}
\crefname{section}{Section}{Sections}
\Crefname{section}{Section}{Sections}
\crefname{table}{Table}{Tables}

\usepackage{flushend}

\graphicspath{Figures/}

\title{Probabilistic forecasting of power system imbalance\\ using neural network-based ensembles}
\author{Jonas Van~Gompel\textsuperscript{a,}\corref{cor1}}
\author{Bert Claessens\textsuperscript{a,b}}
\author{Chris Develder\textsuperscript{a}}
\address{\textsuperscript{a}IDLab, 
Ghent University -- imec, Technologiepark Zwijnaarde, 
Belgium. \\
\textsuperscript{b} Beebop, Belgium}
\cortext[cor1]{Corresponding author.\\ \indent \hspace{4.5pt} E-mail address: \href{mailto:Jonas.VanGompel@UGent.be}{\nolinkurl{Jonas.VanGompel@UGent.be}}
}

\date{\today}

\begin{document}

\begin{frontmatter}

\begin{abstract}
Keeping the balance between electricity generation and consumption is becoming increasingly challenging and costly, mainly due to the rising share of renewables, electric vehicles and heat pumps and electrification of industrial processes. Accurate imbalance forecasts, along with reliable uncertainty estimations, enable transmission system operators (TSOs) to dispatch appropriate reserve volumes, reducing balancing costs. Further, market parties can use these probabilistic forecasts to design strategies that exploit asset flexibility to help balance the grid, generating revenue with known risks.
Despite its importance, literature regarding system imbalance (SI) forecasting is limited. Further, existing methods do not focus on situations with high imbalance magnitude, which are crucial to forecast accurately for both TSOs and market parties.
Hence, we propose an ensemble of C-VSNs, which are our adaptation of variable selection networks (VSNs). 
Each minute, our model predicts the imbalance of the current and upcoming two quarter-hours, along with uncertainty estimations on these forecasts. We evaluate our approach by forecasting the imbalance of Belgium, where high imbalance magnitude is defined as $|$SI$| > 500\,$MW (occurs 1.3\% of the time in Belgium). For high imbalance magnitude situations, our model outperforms the state-of-the-art by 23.4\% (in terms of continuous ranked probability score (CRPS), which evaluates probabilistic forecasts), while also attaining a 6.5\% improvement in overall CRPS. 
Similar improvements are achieved in terms of root-mean-squared error. 
Additionally, we developed a fine-tuning methodology to effectively include new inputs with limited history 
in our model. 
This work was performed in collaboration with Elia (the Belgian TSO) to further improve their imbalance forecasts, 
demonstrating the relevance of our work. 
\end{abstract}


\end{frontmatter}

\section{Introduction}
The dynamics on both the generation and demand side of the power grid are rapidly evolving. On the generation side, countries are transitioning to renewable energy sources to limit carbon emissions and increase energy independence. However, the output of renewables is strongly dependent on weather, making it less predictable than, e.g., a gas turbine~\cite{DSO_flexibility_review}.
On the demand side, the increasing electrification of sectors such as transport (electric vehicles and trains), heating (heat pumps) and industrial processes (e.g., ovens for steel production) amplifies the volatility of electricity consumption~\cite{electrification_nature}.
Consequently, keeping the balance between generation and demand to ensure grid stability is 
increasingly challenging and expensive~\cite{2018_imb_price_forecasting}. 

Transmission system operators (TSOs) are responsible for balancing the grid by activating reserves. 
However, it is costly to both ensure sufficient reserve capacity is available, and to activate them when needed.
To limit these balancing costs, it is imperative that the needs for reserves are not overestimated,
which is complicated by the grid evolutions described above. Therefore, a probabilistic forecaster of the imbalance, which predicts future imbalances and estimates the uncertainty for each forecast,
is an invaluable tool for TSOs~\cite{2019_QRF_SI_forecast}. 
Additionally, European countries are joining the unified market for manual frequency restoration reserves (mFRR), known as the manually activated reserve initiative (MARI), which requires TSOs to submit their mFRR needs 27 minutes before the mFRR is fully activated~\cite{mfrr_EU_rules}. This necessitates accurate imbalance forecasts relatively far into the future.

Probabilistic imbalance forecasts also enable market parties to estimate the expected revenue and risks of imbalance settlement. 
For instance, the probability of activating 
in the wrong direction can be estimated through probabilistic forecasts, allowing market parties to develop asset strategies which generate revenue with minimal risk of reacting in the wrong direction.
Encouraging market parties to actively participate in the imbalance settlement mechanism 
is also advantageous for the TSO, since less reserves need to be activated when market parties are already partly correcting the imbalance. Hence, accurate imbalance forecasts with reliable uncertainty estimations benefit both TSOs and market parties.

Despite the importance of imbalance forecasting, surprisingly few works have tackled this problem~\cite{2019_QRF_SI_forecast, 2021_VSN_RNNSearch_SI_forecast}. 
Although approaches have been proposed to model the area control error (ACE)~\cite{2022_wavelet_LSTM_ACE_forecast} or the grid frequency~\cite{2015_exp_smoothing_freq_forecasting, 2020_freq_forecasting}, 
we focus on forecasting the system imbalance (SI), also known as the open loop ACE. The ACE itself refers to the remaining imbalance after the activated reserve volume has been taken into account, 
while SI is the imbalance as if no reserves were activated. Both the ACE and SI are expressed in megawatts (MW). We focus on forecasting the SI because it determines how much reserves should be activated by the TSO. In turn, the volume of activated reserves determines the imbalance price, which is a key indicator for market parties. In the rest of this work, the term imbalance  refers to SI, not ACE.

The seminal work of Garcia and Kirschen~\cite{2006_NN_SI_forecast} was the first to propose a method for imbalance forecasting, consisting of a multilayer perceptron (MLP) to predict the median imbalance with either daily or four-hourly resolution. These predictions are based on load forecasts and its recent errors, imbalance prices, the volume traded on the intraday market, the total day-ahead (DA) generation nomination and the day of the week.
Contreras additionally includes wind forecasts, temperature, activated reserves and traded volume in the DA market and its price~\cite{2016_RF_SI_forecast_thesis}. These inputs are given to a random forest regressor to forecast the imbalance with hourly resolution. 
Makri et al.\ proposed a recurrent neural network with six Long Short-Term Memory (LSTM) layers to predict the average imbalance over the next 30 minutes, using similar inputs as previous studies~\cite{2021_LSTM_SI_forecasting}.

All of the aforementioned approaches are limited to point forecasts, i.e., the forecasting uncertainty is not modeled. Hence, Salem et al.\ propose a quantile regression forest (QRF) to obtain \emph{probabilistic} forecasts of the imbalance over the next two hours with five minute granularity, based on the imbalance of the past two hours and temporal features~\cite{2019_QRF_SI_forecast}. Since a QRF can only predict a single timestep, 24 separate QRF models need to be trained to forecast the next two hours with five minute granularity. Finally, Toubeau et al.\ developed an attention-based neural network to forecast the imbalance over the next four hours with quarter-hourly resolution, along with an accurate estimate of the forecast uncertainty via quantile regression~\cite{2021_VSN_RNNSearch_SI_forecast}. Specifically, the model consists of the RNNSearch encoder-decoder architecture proposed in~\cite{2014_RNNSearch_paper} stacked on top of the variable selection network (VSN) from~\cite{2021_TFT_paper}. This VSN+RNNSearch model was shown to significantly outperform the QRF approach. The model's input includes the past imbalances, temporal features, load and wind forecasts, cross-border exchanges and DA nominations, along with metering from gas and pumped hydroelectric plants. 

\begin{figure}
    \centering
    \includegraphics[width=\linewidth]{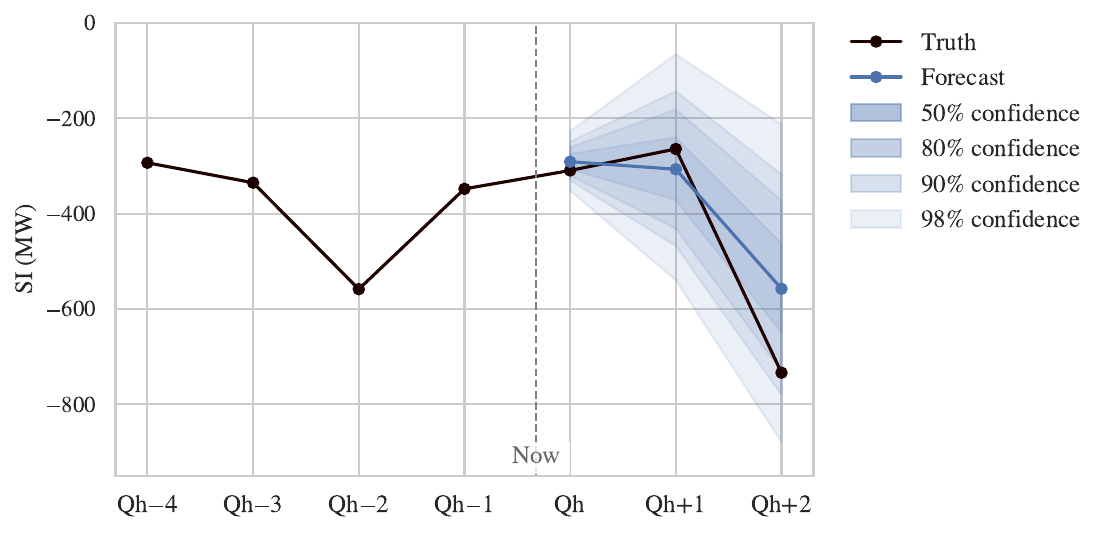}
    \caption{Example of a forecast made by our model, predicting the imbalance of the current and upcoming two quarter-hours (Qh). The solid blue line represents the forecast for the 0.5 quantile, while the other quantile predictions determine the model's confidence by providing prediction intervals.}
    \label{fig:forecast example}
\end{figure}

Both TSOs and market parties are particularly interested in accurate forecasts of spikes in imbalance magnitude, since activation of sufficient reserves is required to ensure grid stability and attractive imbalance prices are likely. Nevertheless, existing methods for imbalance forecasting do not focus on high imbalance magnitude situations. Hence, we propose a neural network-based ensemble for probabilistic imbalance forecasting which considerably outperforms existing models, especially when the imbalance magnitude is high. This is achieved by modifying the VSN architecture 
and by making our model focus on accurately forecasting imbalance spikes through appropriate weighing of its loss function. Additionally, our ensemble is designed to maximize diversity among its submodels, further increasing its performance. The proposed model is detailed in \cref{Proposed model}, of which \cref{fig:forecast example} shows an example of a forecast. In \cref{Performance comparison}, we compare the performance of our model to recently proposed imbalance forecasters~\cite{2019_QRF_SI_forecast, 2021_VSN_RNNSearch_SI_forecast} and the state-of-the-art general-purpose forecasting architecture TSMixer~\cite{2023_TSMixer}, which are described in \cref{Benchmark models}.
The performance impact of each component of our model is quantified in \cref{Ablation of model components}. 

In contrast to existing methods, our model updates its forecast each minute based on the latest information, allowing market parties and TSOs to make up-to-date decisions. In particular, the required mFRR volume needs to be communicated 27 minutes in advance to MARI, which should be based on the latest imbalance forecast. 

While previous works already provide analysis which input features are relevant for imbalance forecasting, we additionally consider the recent imbalance of assets that react to imbalance prices, enabling our model to anticipate these reactions. 
The performance impact of the input features is quantified via an ablation study in \cref{Ablation study of inputs}.

The main contributions of this work are listed below.
\begin{itemize}
    \item We propose an ensemble of C-VSNs, which are an adaptation of variable selection networks (VSNs), to forecast the imbalance of the next three quarter-hours. Our model also provides trustworthy estimates of the forecast uncertainty via quantile regression, which is valuable information for both TSOs and market parties. 
    \item  Our C-VSN ensemble is extensively compared to state-of-the-art models, revealing that both the point forecasts and uncertainty estimations of our model are significantly more accurate than existing approaches, especially when the true imbalance magnitude is high (i.e., $|\text{SI}| > 500\,\text{MW}$ for Belgium). 
    \item To exploit new features which only have a few months of historical data instead of years,
    we developed a fine-tuning approach for our model. We show that fine-tuning on only four months of a new feature already provides over half of the performance gain of the model trained on the full three years of the new feature.
\end{itemize}

\section{Input data} \label{Input data}

An overview of the input features and their horizons is given in \cref{tab:inputs overview}. All data was downloaded from the publicly available Open Data Elia portal~\cite{open_data_elia}, except for the distribution system operator day-ahead nominations (DSO DA nominations) and the asset data, which were obtained directly from Elia. The data ranges from September 1, 2019 until June 30, 2023. 
In the following, we clarify what each feature in \cref{tab:inputs overview} entails, where needed. 

The net regulating volume (NRV) is the net volume of reserves activated by the TSO. The net cross-border nomination is the sum of the nominations for all Belgian borders, including both DA and intraday (ID) nominations.

Regarding the asset data, we include three asset types that react to the imbalance to generate revenue through the imbalance settlement mechanism. To ensure the privacy of the market parties, we do not specify which asset types are provided. We include the metering, DA nominations and activated mFRR volume of each asset type. Metering refers to the measured generation or consumption of the asset type, where negative values denote consumption.
Finally, we include the recent imbalance of each asset type, which is defined as metering – DA nominations – activated mFRR volume. Since the asset imbalances contribute directly to the SI, these are relevant for SI forecasting. To give an idea of the scale of the assets, their combined imbalance ranges between $-1830$\,MW and $1250$\,MW.

\begin{figure}
    \centering
    \includegraphics[width=.8\linewidth]{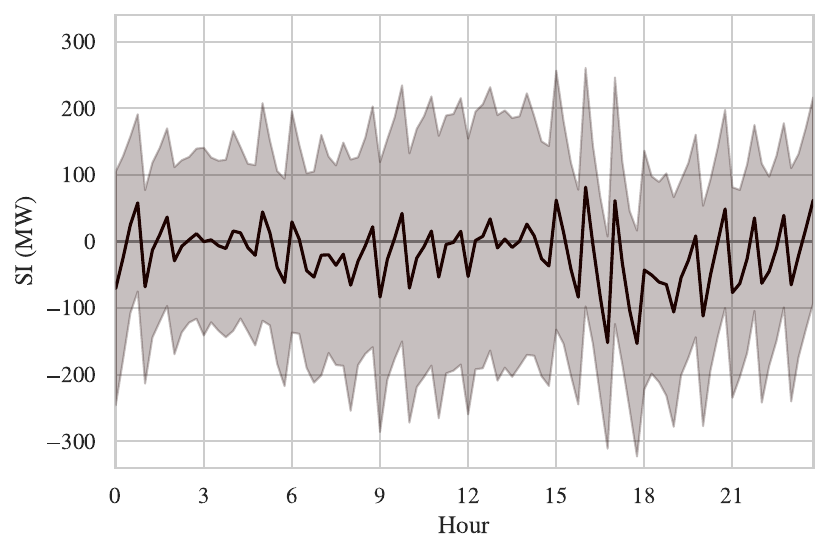}
    \caption{The quarter-hourly average and standard deviation of the Belgian imbalance over the entire dataset. Note that the imbalance average and volatility strongly depends on the time of day.
    }
    \label{fig:average SI over day}
\end{figure}

Time-related inputs are trivial to include, but nevertheless convey important information to the model. For instance, the average and volatility of the Belgian SI varies significantly over a day, as visualized in \cref{fig:average SI over day}. Therefore, we include the Qh of the day, ranging from 1 to 96, and the minute of the hour, ranging from 1 to 60. For neural network-based models, these two features are embedded before further processing by the model (see \cref{C-VSN architecture}). Further, seasonal information is provided via a cosine with yearly period, i.e., $\cos (2\pi t/(365.25\cdot24\cdot60))$ for the $t$-th minute of the data. Since consumption and generation patterns might deviate on holidays, we include a time series with value 1 during a holiday and 0 otherwise. To account for evolutions in the grid dynamics over the period {2019 -- 2023}, we encode the recentness of the data as a linearly increasing value from 0 to 1 over the training set (see \cref{Data preprocessing}). More recent data  (i.e., the validation and test sets) always have recentness set to 1.

The Net DSO DA nomination is the sum of DA nominations of all Belgian DSOs. This value ranges from $860\,$MW to $-9620\,$MW, where positive and negative values indicate generation and consumption, respectively. 

Finally, for each non-temporal feature, we include the change with respect to the previous timestep as additional input:
\begin{equation} \label{eq:delta feature}
    \Delta v_t = v_t - v_{t-1}
\end{equation}
As discussed further in \cref{Ablation of model components}, explicitly including the delta of (non-temporal) features results improves performance and is trivial to implement. 

The total number of input features $N_f$ is 91, where the past and future of, e.g., DSO DA nominations are counted as two features. Specifically, there are 43 non-time features, their deltas and 5 time features, resulting in $N_f=43\cdot2 + 5=91$. The number of input timesteps is $N_c=15$, determining the history and future context length the model has access to. We found that providing more context does not further improve performance. 

\begin{table*}
    \centering
        \begin{tabular}{L{3.2cm}L{5.4cm}C{1.1cm}C{1.1cm}C{1.2cm}C{1.2cm}H}
        \toprule
        Group name & Feature name & Last 15\thinspace Qh & Next 15\thinspace Qh & Last $\text{15\thinspace min}$ & Next 15\thinspace min & Include differenced\\ 
        \midrule
        \multirow{2}{3cm}{SI \& NRV} 
        & SI  & \checkmark & & \checkmark & & \checkmark \\
        & NRV & \checkmark & & \checkmark & & \checkmark \\
        
        \arrayrulecolor{black!30}\midrule 
        \multicolumn{2}{l}{Net cross-border nomination (DA + ID)} & \checkmark & \checkmark & & & \checkmark \\
        
        \arrayrulecolor{black!30}\midrule 
        \multirow{4}{3cm}{Asset data} 
        & Asset imbalance & \checkmark & &  \checkmark & &  \\
        & Asset metering & \checkmark & &  \checkmark & &  \\
        & Asset DA nominations & \checkmark & \checkmark &  & &  \\
        & Asset mFRR activations & \checkmark & & \\
        
        \arrayrulecolor{black!30}\midrule 
        \multirow{5}{3cm}{Time features} 
        & Qh of day ($\in \{1, \dots, 96 \})$ & & \checkmark &  & &  \\
        & Min of hour ($\in \{1, \dots, 60 \})$ & & &  & \checkmark &  \\
        & Year cosine ($\in [0, 1])$ & & \checkmark &  & &  \\
        & Holidays ($\in \{0, 1\})$ & & \checkmark &  & &  \\
        & Recentness ($\in [0, 1])$ & & \checkmark &  & &  \\

        \arrayrulecolor{black!30}\midrule 
        \multirow{2}{3cm}{PV generation} 
        & Measured PV generation & \checkmark & & & & \checkmark \\
        & PV generation forecast & & \checkmark &  & & \checkmark \\

        \arrayrulecolor{black!30}\midrule 
        \multicolumn{2}{l}{Net DSO DA nomination} & \checkmark & \checkmark & & & \checkmark \\
        
        \arrayrulecolor{black!30}\midrule 
        \multirow{2}{3cm}{Load forecasts} 
        & ID load forecast & \checkmark & \checkmark & & & \checkmark \\
        & Recent load forecast & \checkmark & \checkmark & & & \checkmark \\

        \arrayrulecolor{black!30}\midrule 
        \multirow{5}{3cm}{DA generation nominations} 
        & Total DA nomination & & \checkmark & & & \checkmark \\
        & Gas DA nomination & & \checkmark & & & \checkmark \\
        & Nuclear DA nomination & & \checkmark & & & \checkmark \\
        & Hydro DA nomination & & \checkmark & & & \checkmark \\
        & Wind DA nomination & & \checkmark & & & \checkmark \\
        
        \arrayrulecolor{black!30}\midrule 
        \multicolumn{2}{l}{Imbalance price} & \checkmark & & \checkmark & & \checkmark \\

        \arrayrulecolor{black!30}\midrule 
        \multicolumn{2}{l}{Wind generation forecast} & & \checkmark & & & \\

        \arrayrulecolor{black}\bottomrule
        \end{tabular}
    \caption{Overview of the input features and their time horizons. Data regarding the current quarter-hour (Qh) is included in the ``Next 15\thinspace Qh", since it is only known after the current Qh has ended. The groups are sorted in descending order of importance (i.e., their impact on performance, see the ablation study in~\cref{Ablation study of inputs}). The unit of each feature is MW, except for time features and the imbalance price. The latter has unit \EUR/MWh.}
    \label{tab:inputs overview}
\end{table*}

\section{Data preprocessing} \label{Data preprocessing}

Before the data is used to train machine learning models, the following standard preprocessing steps are taken.
\begin{itemize}
    \item The dataset is first divided into training, validation and test sets. The training set ranges from September 2019 to November 2022, the validation set consists of December 2022 and the test set ranges from January 2023 to June 2023. The validation set is used for hyperparameter tuning and early stopping. 
    All results presented in this work are for predictions for the test set.
    \item Since neural networks are sensitive to magnitude differences in their input features, each feature is rescaled to mean 0 and standard deviation 1 over the training set.
    \item Finally, we apply a sliding window to the feature time series to create input samples. Each minute $t$, the window takes the past and/or upcoming $N_c$ values in the time series (which can be either in Qh or min resolution), corresponding to the check marks in \cref{tab:inputs overview}. Each input sample is a $N_c \times N_f$ matrix. The corresponding label $\bm{s}_t$ is a vector containing the imbalance of the current and two upcoming Qh. For example, if a forecast is made at 00:03, the label $\bm{s}_t$ consists of the net imbalance over 00:00 -- 00:15, 00:15 -- 00:30 and 00:30 -- 00:45. 
\end{itemize}

\section{Proposed model} \label{Proposed model}
Here, we present the architecture, loss function, training algorithm and ensembling scheme of the C-VSN ensemble, which is our best performing model (see \cref{Performance comparison}).

\subsection{C-VSN architecture} \label{C-VSN architecture}

\begin{figure*}
    \centering
    \includegraphics[width=.65\linewidth]{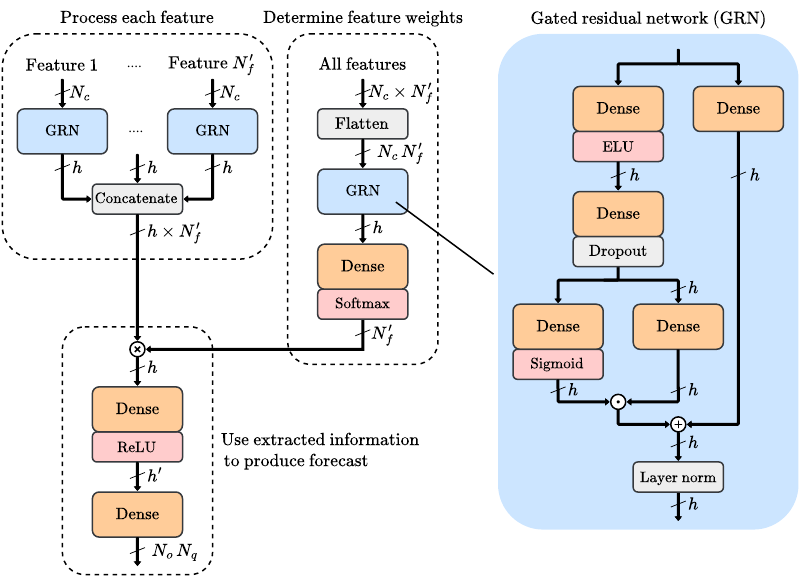}
    \caption{Visualization of the C-VSN model. Next to each arrow, the dimensions of the vector or matrix that is passed is indicated. The~$\otimes$~symbol denotes matrix multiplication, $\odot$ represents element-wise multiplication and $\oplus$ is element-wise addition.}    \label{fig:VSN_diagram}
\end{figure*}

Our model is an adaption of the VSN proposed in~\cite{2021_TFT_paper}. As visualized in \cref{fig:VSN_diagram}, the main steps of the model are:
\begin{enumerate}
    \item Process each of the $N_f$ input features separately using separate gated residual networks (GRNs).
    \item Determine the most relevant features for a sample, again using a GRN.
    \item Weigh the extracted representation of each feature, obtained in step~1, using the feature weights of step~2. Then, use this final representation to produce the forecast.
\end{enumerate}
The combination of using feature weights and GRNs provides significant flexibility in how the model processes each feature, which gives it an edge over the other tested models listed in \cref{Benchmark models}. The feature selection allows the model per instance to suppress certain input features if they are deemed less relevant for a particular forecast. Additionally, the GRNs allow the model to apply non-linear processing only where needed, as discussed in the next paragraph. 
As demonstrated in~\cite{dlinear}, this is specifically relevant for forecasting problems, where linear models with minor adaptations are shown to outperform complex transformer-based models. 
We will now go over each aspect of our model in more detail.

The model does not contain any convolutional, recurrent or attention layers, instead relying entirely on fully-connected layers (denoted as Dense in \cref{fig:VSN_diagram}). The aim of the ReLU and ELU activation functions is to introduce nonlinearities into the model~\cite{2015_ELU}, while the softmax is used to obtain the feature weights. The purpose of the sigmoid activation in the GRN is to act as a gate for the non-linear processing branch of the GRN~\cite{2021_TFT_paper}, which is the GRN's left branch in \cref{fig:VSN_diagram}. If all sigmoid outputs are close to zero, only the linear branch contributes to the output of the GRN. Finally, dropout and layer normalization are introduced in the GRN for regularization.

The main difference between our model and the VSN introduced in~\cite{2021_TFT_paper} is that a VSN processes one input time\-step at a time, while our model processes all timesteps at once, producing a forecast in a single forward pass. This significantly speeds up training and inference while also reaching a higher accuracy, which will be studied in \cref{Ablation of model components}.
Note that the feature weights of the original VSN vary per timestep, while our model provides one set of feature weights per sample.  
Since the feature weights remain constant over the timesteps, we refer to our VSN adaptation as Constant-VSN or C-VSN.

To effectively exploit the time features `Qh of day' and `Min of hour', these are first passed through an embedding layer, mapping each value of the time feature to a five-dimensional vector. Since these embeddings are trained along with the other parameters of the model, time values with similar properties (such as a similar average imbalance in \cref{fig:average SI over day}) can be mapped close to each other in the embedded space~\cite{2013_word2vec}. This enables the model to effectively extract useful information from these time features. Due to the embedding, the number of features for C-VSN is $N'_f=N_f-2+10=99$. 

The other dimensions in \cref{fig:VSN_diagram} are the number of input timesteps $N_c=15$ and output timesteps $N_o=3$, as we are forecasting the current and next two Qh. For each of these Qh, the forecast uncertainty is estimate via quantile regression. As discussed in \cref{Loss function}, we predict nine quantiles, thus $N_q=9$. The hidden dimensions $h$ and $h'$ determine the number of parameters of the internal fully-connected layers and are set to $h = 10$ and $h' = 32$. Higher values cause overfitting, while lower values do not provide sufficient capacity for the model to fully learn the task at hand. This was observed by testing lower and higher values while monitoring the performance on the validation set.

\subsection{Loss function and training algorithm} \label{Loss function}

To enable uncertainty estimations, the model's parameters are optimized by minimizing the quantile loss $\mathcal{L}_Q$. We predict the quantiles $Q=\{0.01, 0.05, 0.1, 0.25, 0.5, 0.75,$ $0.9, 0.95, 0.99\}$, where the true imbalance value should fall below the 0.01 quantile prediction in 1\% of the cases, and similarly for the other quantiles.
The 0.5 quantile predictions are of course used as the point forecasts of the model.

Quantile regression is commonly adopted because it does not make assumptions regarding the underlying distribution of forecasting errors~\cite{2021_VSN_RNNSearch_SI_forecast}. 
A downside of quantile regression is the crossing problem, where the predicted value for quantile $q$ can be lower than the prediction for quantile $q'$, even though $q>q'$ (or vice versa). Although this can be avoided by including a penalty term in the loss or imposing restrictions in the model~\cite{2020_deep_NN_quantile_regression}, these were not implemented since no quantile crossing was observed in our forecasts.

The quantile loss function is defined as 
\begin{align*} \label{quantile loss}
    &\mathcal{L}_Q = \frac{1}{T N_o N_q}\sum_{t=1}^{T} \sum_{i=1}^{N_o} w(s_t^i) \sum_{q\in Q} L_q (s_t^i, \hat{s}_{t,q}^i)  \\ 
    & \text{with} \quad L_q (s_t^i, \hat{s}_{t,q}^i) = 
    \begin{cases}
        q |s_t^i - \hat{s}_{t,q}^i| & \text{if } s_t^i > \hat{s}_{t,q}^i \, ,\\
        (1-q) |s_t^i - \hat{s}_{t,q}^i| & \text{otherwise.}
    \end{cases}
\end{align*}
Here, the first sum runs over all training samples, the second sum over the three Qh ($N_o=3$) that are forecasted, and the third sum over the predicted quantiles. Hence, $q$ takes on the values 0.01, 0.05, etc. An element of the label $\bm s_t$ is denoted as $s_t^i$, meaning $s_t^1$, $s_t^2$ and $s_t^3$ correspond to the true imbalance of the current Qh, Qh+1 and Qh+2, respectively. An output of the model is denoted as $\hat{s}_{t,q}^i$, which is the model prediction for Qh $i$ and quantile $q$. 

The function $L_q$ is simply a weighed version of the mean absolute error (MAE), and is equivalent to the MAE (up to a constant factor $\frac{1}{2}$) for the quantile $q=0.5$. For the 0.01 quantile, $L_q$ will weigh the MAE by a factor $0.01$ if the real imbalance is higher than the model output, while it will weigh the MAE with $0.99$ if the real imbalance is lower. Therefore, it is intuitively clear that the model will learn to predict the $0.01$ quantile with an output trained to minimize this loss.

\begin{figure}
    \centering
    \includegraphics[width=.65\linewidth]{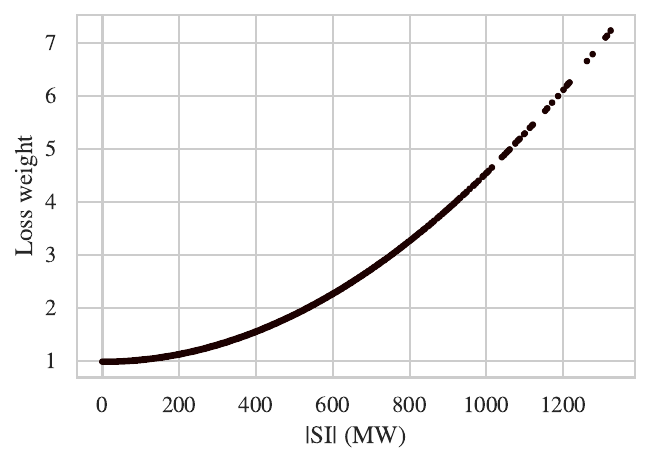}
    \caption{Value of the loss weights of the training labels for $c=0.1$ in \cref{loss weights}.
    }
    \label{fig:loss_weight}
\end{figure}

The weight $w(s_t^i)$ is included in the loss function to prioritize reducing the forecasting errors for high imbalance magnitude situations. Specifically, the weight for label $s_t^i$ is given by
\begin{equation} \label{loss weights}
    w(s_t^i) = 1+ c\cdot\left(s_t^i\right)^2 \, ,
\end{equation}
where $c$ is a constant determining how strongly cases of high imbalance magnitude should be weighed. Note that the labels $s_t^i$ are rescaled to mean 0 and standard deviation 1, and are thus not imbalance values in MW. As described further in \cref{Ensembling}, the value of $c$ is different for each ensemble submodel and ranges from 0 to 0.1. Higher values of $c$ are not advised because the trade-off between accuracy for low and high $|$SI$|$ becomes less interesting, i.e., the performance for low $|$SI$|$ drops too much. The quadratic form of \cref{loss weights} was chosen empirically, as the models trained with these weights show a better trade-off between accuracy for low and high $|$SI$|$ than models trained with weights that scale linearly, cubically or exponentially with $|$SI$|$.
The value of $w$ for $c=0.1$ is depicted in \cref{fig:loss_weight}, which shows that labels of the most extreme imbalance values are weighed at most 7.2 times more in the loss function than labels of {SI $\approx 0$}.

The model is trained for 10 epochs using the Adam optimizer~\cite{adam} with learning rate 0.001 and batch size 128. To avoid overfitting, we use early stopping, where training is stopped prematurely if the evaluation metrics (listed in \cref{Evaluation metrics}) on the validation set no longer improve.

\subsection{Ensembling} \label{Ensembling}

A simple but effective way to improve the model's performance is to create an ensemble, where the predictions of multiple C-VSNs, trained with different initial parameters, are averaged. Although this standard ensemble already reduces overfitting, encouraging diversity between the ensemble's submodels 
can further improve performance~\cite{2019_ensemble_diversity_good}. For this purpose, we vary the following aspects per submodel.
\begin{itemize}
    \item As in a standard ensemble, a different seed is used for the parameter initialization of each submodel.
    \item Bootstrapping: each submodel is trained with $T$ samples randomly drawn with replacement from the training set.
    \item Submodels 1, 3, 5, etc.\ predict SI, whereas submodels 2, 4, 6, etc.\ predict $\Delta$SI, where $\Delta$SI denotes the change in SI compared to the previous Qh.
    \item Increase the loss weighing constant $c$ in \cref{loss weights} from 0 to 0.1 in steps of 0.005. This means that submodel 1 is trained with $c=0$, submodel 2 with $c=0.005$, \dots, submodel 21 with $c=0.1$.
\end{itemize}
The performance impact of each point is discussed in \cref{Ablation of model components}. 
We found that additional performance gain from using more than 21 submodels is negligible.

\section{Reference models} \label{Benchmark models}

To thoroughly evaluate our model, we compare its performance to five other models, two of which are proposed specifically for probabilistic imbalance forecasting \cite{2019_QRF_SI_forecast, 2021_VSN_RNNSearch_SI_forecast}. The reference models are described below. 
\begin{itemize}
    \item \textit{Linear}: a quantile linear regression model, implemented as a single fully-connected layer.
	\item QRF: a quantile regression forest, as proposed in~\cite{2019_QRF_SI_forecast}. We use the same hyperparameter values as~\cite{2019_QRF_SI_forecast}: the forest consists of 100 trees with a minimum leaf size of 10 and mean squared error as splitting criterion.
	\item \textit{GradBoost}: a gradient boosting regression tree, which was also used as baseline in~\cite{2021_VSN_RNNSearch_SI_forecast}. Following~\cite{2021_VSN_RNNSearch_SI_forecast}, the number of boosting iteration is set to 100 with learning rate 0.1. Note that $N_o N_q$ separate GradBoost models are trained, since each model only forecasts one timestep and one quantile.
    \item \textit{VSN+RNNSearch}: the model proposed in~\cite{2021_VSN_RNNSearch_SI_forecast}, consisting of an RNNSearch encoder-decoder architecture stacked on top of a VSN. Using the same  hyperparameter choices, the hidden dimension of the VSN and encoder bidirectional LSTMs is 24, while the hidden dimension the decoder bidirectional LSTM is 12.
    \item \textit{TSMixer}: the recently proposed Time-Series Mixer architecture~\cite{2023_TSMixer}, which consecutively applies MLPs to both the time and feature dimensions and achieves state-of-the-art results on common forecasting benchmarks. In our experiment, we use a single time and feature MLP block with hidden dimension 32 and layer normalization.
\end{itemize}
To ensure fair comparison, all models use the same input data and preprocessing steps, as previously detailed in \cref{Input data} and \cref{Data preprocessing}. Note that QRF and GradBoost cannot embed the time feature `Qh of day' and `Min of hour', since tree-based models are not trained using backpropagation. Further, the input samples are flattened from shape $N_c \times N_f$ to $N_c N_f$ before they are fed to the linear, QRF and GradBoost models, as these models do not support a time dimension.

\section{Results and discussion}

\subsection{Evaluation metrics} \label{Evaluation metrics}

\begin{figure*}
    \centering
    \includegraphics[width=\linewidth]{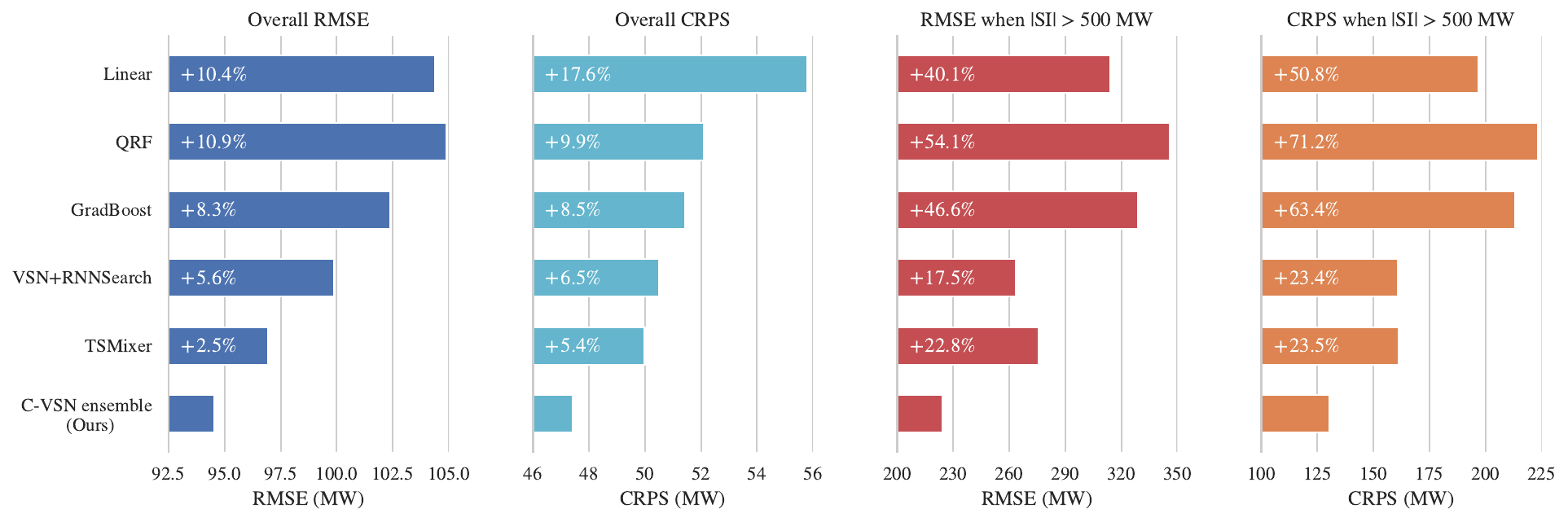}
    \caption{The results achieved by each model on the test set, in terms of the overall RMSE, the overall CRPS, and the RMSE and CRPS considering only instances where the true imbalance magnitude is larger than $500\,$MW. The percentages denote the increase in RMSE or CRPS compared to our C-VSN ensemble.}
    \label{fig:all_metrics}
\end{figure*}

To assess the accuracy of the models' point forecasts (i.e., the predictions for the 0.5 quantile), we determine the root mean squared error (RMSE) on the test set. 

\begin{figure}
    \centering
    \includegraphics[width=.85\linewidth]{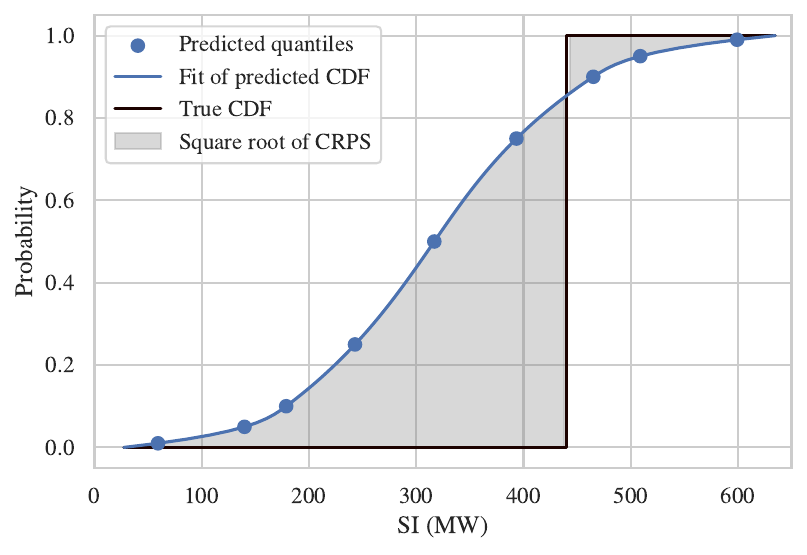}
    \caption{Example of how the CRPS is determined. The blue dots are the forecasted quantiles for Qh+2 in \cref{fig:forecast example}. The blue line is the quadratic spline fit to the quantile predictions, representing the predicted CDF, while the black line is the true CDF (i.e., jumping to 1 at the true value of the SI on Qh+2). The overall CRPS is obtained by averaging the square of the shaded area over all forecasts in the test set.}
    \label{fig:CRPS_visualization}
\end{figure}

The full probabilistic forecasts are evaluated via the continuous ranked probability score (CRPS)~\cite{2007_CRPS}, which is visualized for an exemplary prediction in \cref{fig:CRPS_visualization}. The CRPS is defined as
\begin{equation} \label{eq:CRPS}
    \text{CRPS}\,(F_t^i, s_t^i) = \int_{-\infty}^{\infty}\Big(F_t^i(x) - H(x-s_t^i)\Big)^2\,\mathrm{d}x\,,
\end{equation}
where $F_t^i$ is the forecasted cumulative distribution function (CDF), $s_t^i$ is the true imbalance value and $H$ is the Heaviside step function, representing the CDF of the true imbalance (black line in \cref{fig:CRPS_visualization}). Since quantile regression produces forecasts of discrete quantiles, and not a continuous CDF, we fit a cubic spline to the quantile predictions to obtain the forecasted CDF (blue line in \cref{fig:CRPS_visualization}). Then, we numerically calculate the integral in \cref{eq:CRPS}. Finally, the CRPS is averaged over all test set forecasts.
Similar to RMSE, lower values of CRPS are better. Clearly, the CRPS is small when the predicted quantiles are closely centered around the true imbalance. 
When applied to deterministic forecasts, i.e., when the forecasted CDF is also a Heaviside function, the CRPS reduces to the mean absolute error~\cite{2007_CRPS}.

Considering that it is particularly important to anticipate situations of high imbalance magnitude, we also monitor the RMSE and CRPS taking only the instances of high imbalance magnitude into account, defined as $|$SI$|>500$\,MW. This corresponds to roughly $1.3\%$ of the Qh in the test set. Thus, the impact of high imbalance magnitude situations on the overall RMSE and CRPS is limited, which is why it is useful to also separately monitor the metrics for $|$SI$|>500$\,MW situations.

\subsection{Performance comparison} \label{Performance comparison}

\Cref{fig:all_metrics} presents the test set results of all models. 
The notable performance gap between the linear and tree-based models on the one hand and neural network-based models on the other hand indicate that the latter are well-suited for capturing the intricate and non-linear dynamics of the electricity market.
Further, our C-VSN ensemble outperforms all other models, including the QRF and VSN+RNNSearch models specifically proposed for imbalance forecasting. Especially the performance for high imbalance magnitude is striking, outperforming the second-best model by 17.5\% for RMSE and 23.4\% for CRPS.
Although TSMixer also performs well, it is still not as accurate as C-VSN ensemble, even when an ensemble of TSMixers is used. 

Note that our results also confirm the findings in~\cite{2021_VSN_RNNSearch_SI_forecast}, namely that the overall CRPS of QRF 
is roughly 3.4\% higher than that of VSN+RNNSearch. 
In turn, the C-VSN ensemble outperforms VSN+RNNSearch in terms of overall CRPS by 6.5\%.

 \begin{figure*}
    \centering
    \includegraphics[width=\linewidth]{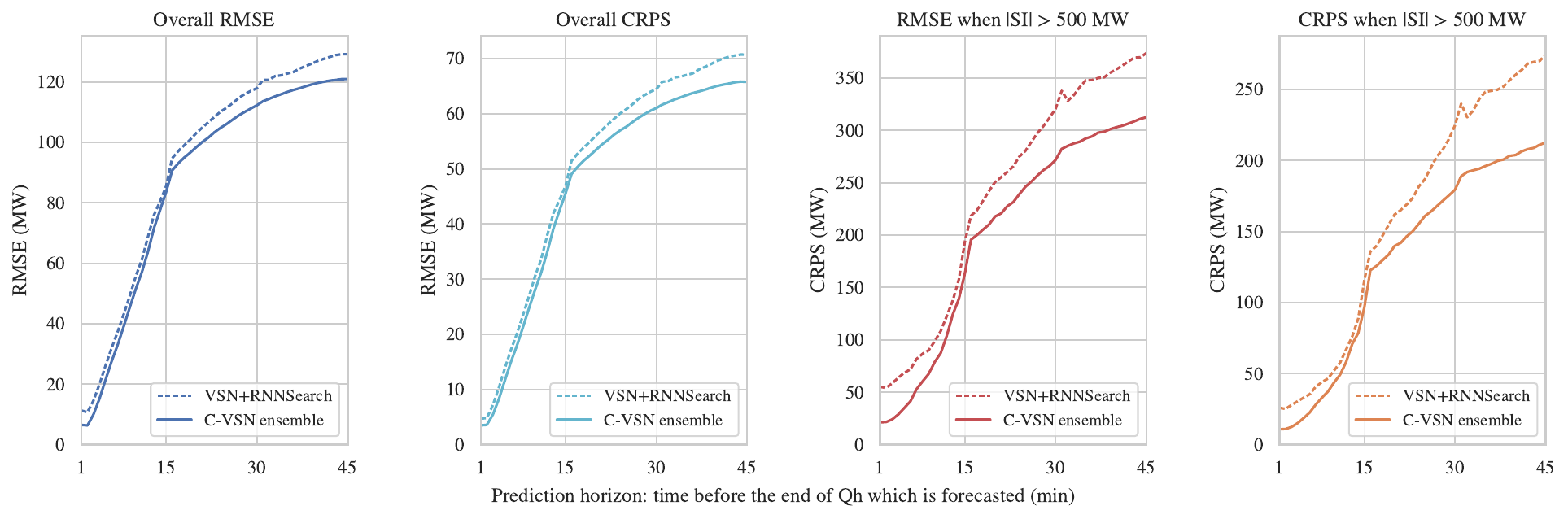}
    \caption{The evolution of the metrics depending on how far into the future is being forecasted. Naturally, the predictions made 1--15\thinspace min into the future are for the current Qh, 16--30 are for Qh+1 and 31--45 are for Qh+2, each produced by their respective output. 
    }
    \label{fig:metrics over 45 mins}
\end{figure*}

Since each minute a new prediction is made for the current and upcoming two Qh, the accuracy evolves depending on how far into the future is being forecasted. Specifically, as more information becomes available during the Qh, such as the minute-based imbalance values, the accuracy of forecasts made in the 14th minute of a Qh will generally be higher than forecasts made in the first minute. This evolution is presented for the entire forecast horizon in \cref{fig:metrics over 45 mins}. These results show that the performance improvement of our model is most prominent for predictions of Qh+1 and Qh+2. 
This is especially relevant for estimating the TSO's mFRR needs, which must be communicated to the MARI platform 27 minutes in advance. 
The high slope for the current Qh forecasts in \cref{fig:metrics over 45 mins} is due to the strong reliance on the minute-based imbalance, which is of course the most important feature for predicting the current Qh imbalance.
\Cref{fig:metrics over 45 mins} also shows the results for VSN+RNNSearch, which is the best performing model reported specifically for imbalance forecasting. For clarity, the results of the other models are not shown.

\begin{figure}
    \centering
    \includegraphics[width=.9\linewidth]{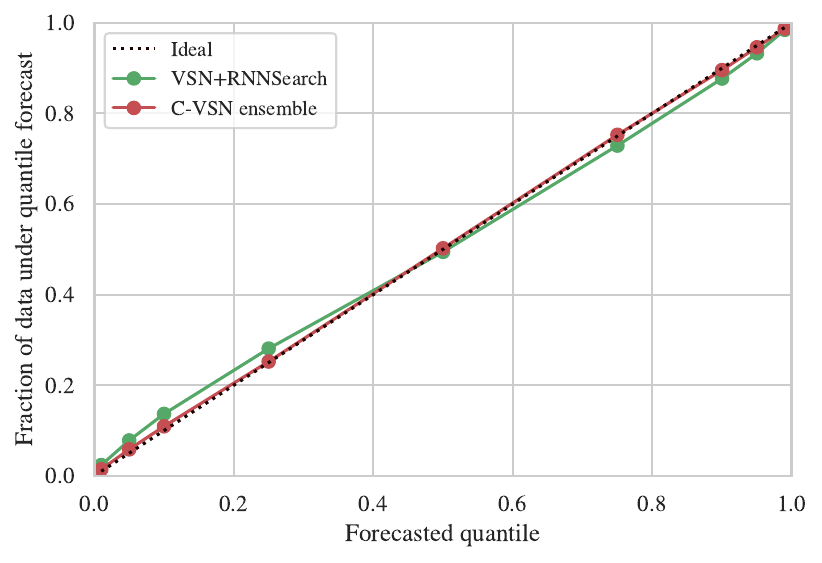}
    \caption{Fraction of the true imbalance values in the test set which are lower than the forecasts for a quantile. Each dot corresponds to one of the forecasted quantiles. Ideally, the fraction should be 0.01 for the 0.01 quantile, and likewise for the other quantiles.}
    \label{fig:fraction_under_quantile}
\end{figure}

To analyze the reliability of the uncertainty estimations of C-VSN ensemble, we determine the fraction of test set labels which are lower than the predictions for a quantile. Ideally, 1\% of the imbalance values should be lower than the 0.01 quantile forecast, 10\% for the 0.1 quantile, and so on. As shown in \cref{fig:fraction_under_quantile}, the fraction of test labels under the quantile forecasts is remarkably close to the ideal values. Such reliable uncertainty estimations are a valuable tool for both TSO reserve dispatching and market parties. 
For instance, market parties can use our model to estimate the probability of balancing in the wrong direction, which is the main risk when participating in the imbalance settlement mechanism.

Although the results for VSN+RNNSearch are also close to the ideal values, the quantile predictions are slightly too high for quantiles $< 0.5$ and too low for quantiles $> 0.5$. This indicates VSN+RNNSearch is slightly overconfident in its forecasts, resulting in prediction intervals that are too narrow.

\subsection{Ablation of C-VSN ensemble components} \label{Ablation of model components}

\begin{figure}
    \centering
    \includegraphics[width=\linewidth]{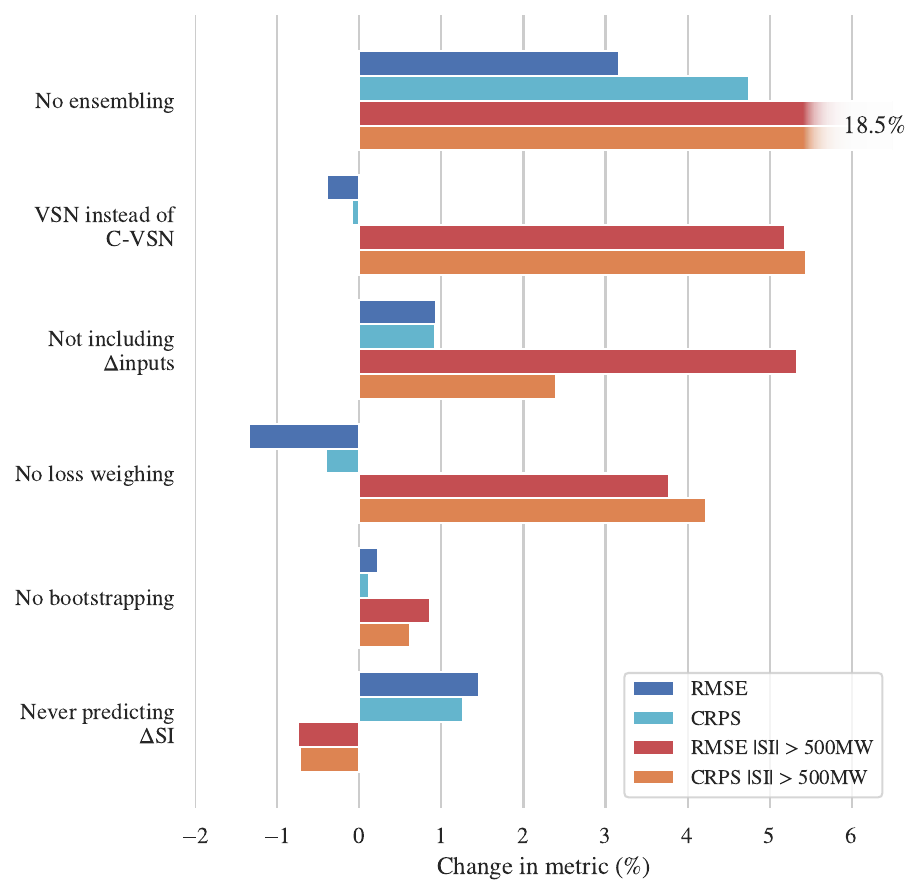}
    \caption{Change in performance of the C-VSN ensemble when removing a component from the model. Note that a positive change indicates the model performs worse when the component is removed.}
    \label{fig:model component ablation}
\end{figure}

To quantify to which degree the various components of our C-VSN ensemble contribute to its performance, we determine the change in performance when a component is removed. The ablation results in \cref{fig:model component ablation} reveal that ensembling significantly benefits the performance. Note that removing ensembling also includes the aspects that encourage diversity between submodels (i.e., loss weighing, alternating between predicting SI and $\Delta$SI, and bootstrapping). When removing only the loss weighing (by setting $c=0$ in \cref{loss weights} for all submodels), the overall RMSE and CRPS slightly improve. This is not surprising, since the loss weights make the model focus on high $|$SI$|$ cases at the expense of low $|$SI$|$ situations, which are much more common. However, we believe the $\sim$\,4\% improvement for high $|$SI$|$ situations justifies this trade-off, considering the real-world relevance of accurately forecasting these cases. Both predicting $\Delta$SI and bootstrapping improve the model's performance by promoting submodel diversity, albeit with limited impact ($\sim$\,1\%).

Despite its simplicity, including the change in input features (see \cref{eq:delta feature}) as additional inputs substantially improves performance, likely because the correlation magnitude between imbalance and $\Delta$feature is often higher than for the feature itself. For instance, the correlation with imbalance is $-0.06$ for the net cross-border nominations and $0.21$ for the change in net cross-border nominations. 

Finally, using the C-VSN adaptation instead of VSN not only speeds up training\footnote{Training performed using an Intel i7-9700 CPU with 16 GB RAM.} of a submodel from 142 minutes to 66 minutes, but also increases the performance for high $|$SI$|$. This can be explained by the fact that VSN produces an output per timestep, which is required when stacking recurrent and/or attention layers on top of the VSN, like in the temporal fusion transformer from~\cite{2021_TFT_paper} or VSN+RNNSearch from~\cite{2021_VSN_RNNSearch_SI_forecast}. However, we observed that the additional complexity from recurrent and attention layers do not contribute towards more accurate imbalance forecasts. Without these layers, the VSN output must be flattened from $N_c\times h$ to $N_c h$ before passing it to the final fully-connected layers, discarding valuable temporal information in the process. Instead, our C-VSN adaptation already processes the time dimension in the per-feature GRNs in \cref{fig:VSN_diagram}, resulting in an output of shape $h$ which is passed to the final fully-connected layers. By processing the temporal dimension more effectively, our C-VSN adaptation outperforms VSN for imbalance forecasting, as shown in \cref{fig:model component ablation}.

\subsection{Ablation of inputs} \label{Ablation study of inputs}

\begin{figure}
    \centering
    \includegraphics[width=\linewidth]{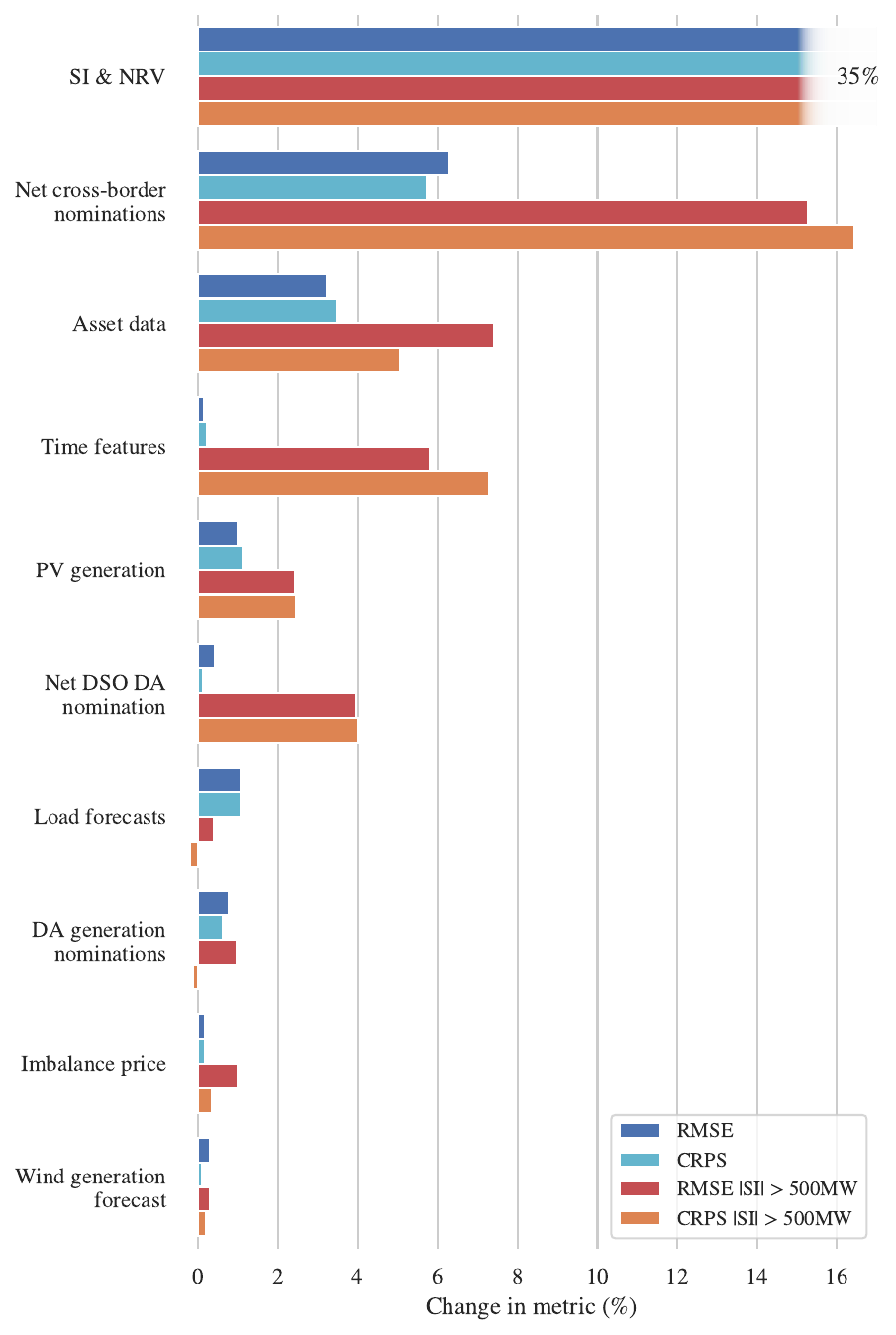}
    \caption{Change in performance of the C-VSN ensemble when trained without an input feature group. The groups are specified in \cref{tab:inputs overview}.}
    \label{fig:ablation results}
\end{figure}

The importance of each feature group is determined by another ablation study, where we monitor the change in performance when the C-VSN ensemble is trained without an input feature group. This feature importance is useful to decide which inputs should be prioritized when developing an imbalance forecaster. The results are presented in \cref{fig:ablation results}, where a higher increase in metrics implies a more important feature. 

Unsurprisingly, the past SI and NRV observations are the most important features. More interesting is that the net cross-border nominations, asset data and even time features have a significant impact on the model's performance. The cross-border nominations determine how much volume will be imported from / exported to neighbouring countries in the future, providing valuable information for imbalance forecasting. 
The asset data is useful because (a) the imbalance of the assets directly contributes to the SI and (b) the assets situationally participate in the imbalance settlement mechanism, 
reducing the imbalance magnitude and generating revenue for the asset owners. 
For instance, when circumstances allow and SI $\ll 0$ (i.e., there is a significant energy shortage in Belgium), industrial processes can be delayed to reduce energy consumption. Conversely, when SI $\gg 0$, power plants can curtail generation to reduce the SI.
Hence, explicitly providing asset data enables the model to anticipate these events and their impact on the SI. 
As previously noted in \cref{fig:average SI over day}, time features also have a significant impact on the imbalance, making them useful for the model. While the impact of the other features is less prominent, all of them still benefit the overall performance.

As a side note, we have also experimented with including information from the Belgian ID market EPEX Spot as input. Specifically, we included the net volume of recently submitted open bids for each of the upcoming Qhs, since this indicates the net imbalance of the portfolios of market parties for the near future, which will contribute to the overall SI. Although there is a correlation of 0.11 between the SI of the next Qh and the net volume of unmatched bids submitted in the past hour, including this feature in the input of our model (or any of the reference models) does not impact performance.

\begin{figure}
    \centering
    \includegraphics[width=\linewidth]{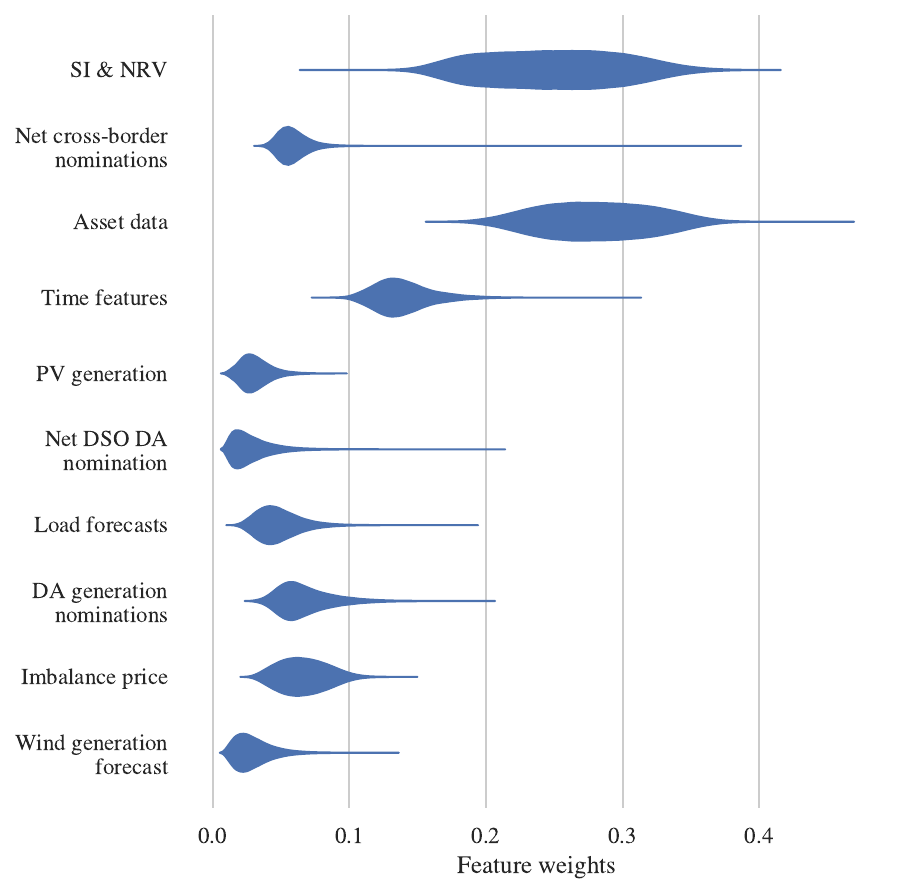}
    \caption{Distribution of the feature weights of C-VSN over the test set forecasts. For clarity, the weights are summed per feature group.}
    \label{fig:Attention weights}
\end{figure}

As described in \cref{C-VSN architecture}, our C-VSN architecture infers suitable feature weights each time a forecast is produced. The distribution of the feature weights are visualized in \cref{fig:Attention weights}. 
As intuitively expected, groups which receive a higher weight are those that have a larger impact on performance. 
These feature weights thus enable some degree of explainability, indicating which inputs had the most significant impact on a particular forecast.

\subsection{Fine-tuning on features with less history}

\begin{figure}
    \centering
    \includegraphics[width=.7\linewidth]{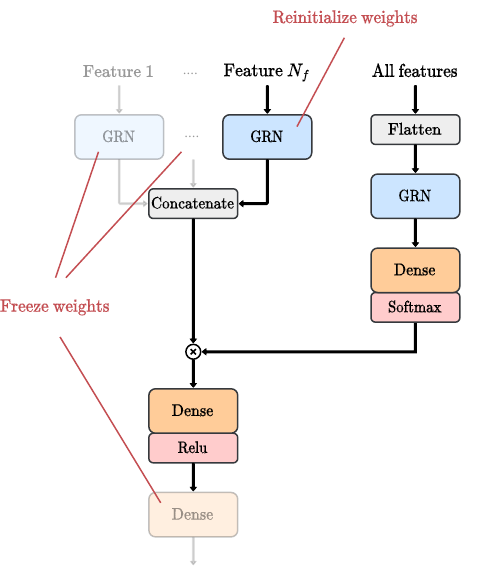}
    \caption{Modifications to the model during fine-tuning on a new feature with less history. In this example, the model is fine-tuned on a single feature $N_f$, which has limited history.}    \label{fig:finetune_diagram}
\end{figure}

In practice, new features may become available that are helpful for imbalance forecasting, but do not yet have a history of multiple years and can therefore not simply be included in the training of the model. Such a feature could be, e.g., the metering measurements of a newly installed asset. 
The proposed C-VSN architecture easily supports incorporating features with limited history by fine-tuning on the recent period for which the new feature is available, e.g., the last few months.
The fine-tuning methodology is visualized in \cref{fig:finetune_diagram} and consists of 4 steps:
\begin{enumerate}
    \item Train the model on the entire training set, where the values of the features with limited history are always set to 0. The training is performed as usual (see \cref{Loss function}).
    \item Reinitialize the parameters of the GRNs corresponding to the features with less history. Further, the state of the Adam optimizer is also reset.
    \item Freeze the parameters of the final fully-connected layer and of the GRNs corresponding to features with full history. Freezing means that the parameters of these layers will no longer be trained, and hence their values will remain fixed.
    \item Train the model on the last month(s) using all features, including the ones with limited history. The training is also performed as described in \cref{Loss function}, except that the number of epochs is set to 1 to avoid overfitting on the small training set.
\end{enumerate}
Steps~2 and~3 are optional, but are beneficial to the performance of the fine-tuned model. The intuition behind step~2 is that during step~1, the model has learned that the features with less history are useless (i.e., always zero), which could lead to undesirable values for the parameters of the corresponding GRNs. Therefore, it is useful to reinitialize these GRNs before fine-tuning. Step~3 is implemented to avoid overfitting on the small training set used in step~4. Specifically, the GRNs of the regular features were already trained on the full history, so retraining these layers on only the last month(s) could cause them to `forget' useful aspects learned in step~1, leading to overfitting. Likewise, the output layer of the model is also frozen to avoid overfitting, because the task at hand (i.e., quantile regression of imbalance) has not changed.

\begin{figure}
    \centering
    \includegraphics[width=\linewidth]{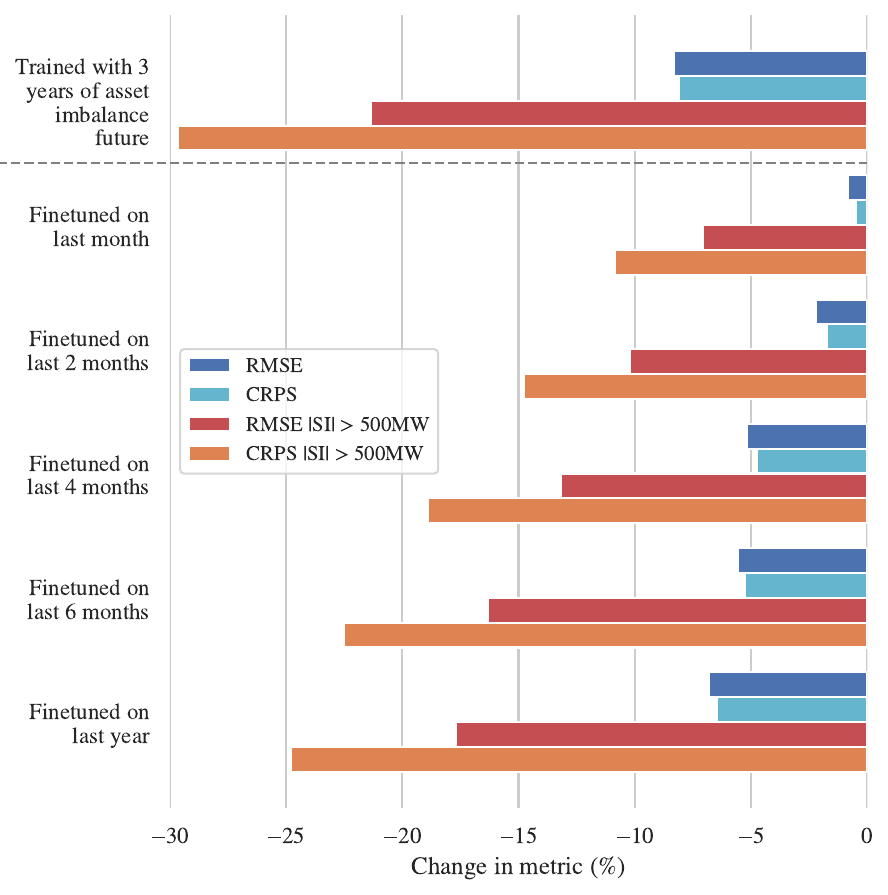}
    \caption{Performance improvement from fine-tuning on the future imbalance of an asset (which is used as a proxy for a new useful feature with limited history). Fine-tuning on only four months already provides over half of the performance gain of the model trained on the full three years of the asset's future imbalance.}    \label{fig:finetune_results}
\end{figure}

We test our fine-tuning method using an asset's imbalance values for the next 15\thinspace Qh as a new feature with limited history. Of course, this is an artificial setup, since future values cannot be used in practice. The aim of this artificial setup is to evaluate whether our model can learn to exploit a new feature with limited history via fine-tuning. 
Unsurprisingly, \cref{fig:finetune_results} shows that a model trained on the full training set with the asset's future imbalance as additional input performs significantly better than a model trained without the asset's future. We will refer to this performance gain as the maximal performance improvement from the asset's future imbalance, since the full 3 years of that data is used.
More importantly, when the model is initially trained without the asset's future imbalance and then fine-tuned using only the last four months, it already captures over half of the maximal performance improvement achievable through the incorporation of the new feature. As expected, more of the maximal accuracy improvement is captured when using longer histories for fine-tuning, as shown in \cref{fig:finetune_results}.

\section{Conclusions}
We propose an accurate model for probabilistic forecasting of the power system imbalance, which can reduce balancing costs for TSOs by predicting reserve requirements. Further, market parties can use our model's probabilistic imbalance forecasts to exploit asset flexibility, generating revenue with known risks while helping to balance the grid. 
In contrast to methods from literature, we focus on accurately forecasting imbalance spikes (both in positive and negative direction), which are particularly relevant for both TSOs and market parties. The performance of our model is extensively compared to state-of-the-art approaches. Additionally, ablation studies are presented to reveal which model components and input features contribute most to our model's performance. Finally, we propose a fine-tuning approach to easily integrate new input features with only a few months of historical data, enabling our model to exploit new information as soon as possible.

\section*{Acknowledgements}
We would like to thank Elia for providing valuable insights and data.


\bibliographystyle{ieeetr}
\bibliography{references}

\begin{thebibliography}{10}

\bibitem{DSO_flexibility_review}
X.~Jin, Q.~Wu, and H.~Jia, ``Local flexibility markets: Literature review on concepts, models and clearing methods,'' {\em Applied Energy}, vol.~261, p.~114387, 2020.

\bibitem{electrification_nature}
G.~Luderer, S.~Madeddu, L.~Merfort, F.~Ueckerdt, M.~Pehl, R.~Pietzcker, M.~Rottoli, F.~Schreyer, N.~Bauer, L.~Baumstark, {\em et~al.}, ``Impact of declining renewable energy costs on electrification in low-emission scenarios,'' {\em Nature Energy}, vol.~7, no.~1, pp.~32--42, 2022.

\bibitem{2018_imb_price_forecasting}
J.~Lago, F.~De~Ridder, and B.~De~Schutter, ``Forecasting spot electricity prices: Deep learning approaches and empirical comparison of traditional algorithms,'' {\em Applied Energy}, vol.~221, pp.~386--405, 2018.

\bibitem{2019_QRF_SI_forecast}
T.~S. Salem, K.~Kathuria, H.~Ramampiaro, and H.~Langseth, ``Forecasting intra-hour imbalances in electric power systems,'' in {\em Proceedings of the AAAI Conference on Artificial Intelligence}, vol.~33, pp.~9595--9600, 2019.

\bibitem{mfrr_EU_rules}
{European Union Agency for the Cooperation of Energy Regulators (ACER)}, ``Implementation framework for the {European} platform for the exchange of balancing energy from frequency restoration reserves with manual activation,'' 2020.
\newblock \url{https://bit.ly/2TPM5Tk}.

\bibitem{2021_VSN_RNNSearch_SI_forecast}
J.-F. Toubeau, J.~Bottieau, Y.~Wang, and F.~Vall{\'e}e, ``Interpretable probabilistic forecasting of imbalances in renewable-dominated electricity systems,'' {\em IEEE Transactions on Sustainable Energy}, vol.~13, no.~2, pp.~1267--1277, 2021.

\bibitem{2022_wavelet_LSTM_ACE_forecast}
H.~Abdeltawab and A.~Radwan, ``Area control error forecasting using deep learning for an interconnected power system,'' in {\em IEEE Power and Energy Conference at Illinois (PECI)}, pp.~1--5, IEEE, 2022.

\bibitem{2015_exp_smoothing_freq_forecasting}
J.~W. Taylor and M.~B. Roberts, ``Forecasting frequency-corrected electricity demand to support frequency control,'' {\em IEEE Transactions on Power Systems}, vol.~31, no.~3, pp.~1925--1932, 2015.

\bibitem{2020_freq_forecasting}
O.~Yurdakul, F.~Eser, F.~Sivrikaya, and S.~Albayrak, ``Very short-term power system frequency forecasting,'' {\em IEEE Access}, vol.~8, pp.~141234--141245, 2020.

\bibitem{2006_NN_SI_forecast}
M.~P. Garcia and D.~S. Kirschen, ``Forecasting system imbalance volumes in competitive electricity markets,'' {\em IEEE Transactions on Power Systems}, vol.~21, no.~1, pp.~240--248, 2006.

\bibitem{2016_RF_SI_forecast_thesis}
C.~Contreras, ``System imbalance forecasting and short-term bidding strategy to minimize imbalance costs of transacting in the spanish electricity market,''
\newblock MSc. Thesis, Comillas Pontifical Univ., 2016.

\bibitem{2021_LSTM_SI_forecasting}
E.~Makri, I.~Koskinas, A.~C. Tsolakis, D.~Ioannidis, and D.~Tzovaras, ``Short-term net imbalance volume forecasting through machine and deep learning: a {UK} case study,'' in {\em IFIP International Conference on Artificial Intelligence Applications and Innovations}, pp.~377--389, Springer, 2021.

\bibitem{2014_RNNSearch_paper}
D.~Bahdanau, K.~H. Cho, and Y.~Bengio, ``Neural machine translation by jointly learning to align and translate,'' in {\em 3rd International Conference on Learning Representations, ICLR 2015}, 2015.

\bibitem{2021_TFT_paper}
B.~Lim, S.~{\"O}. Ar{\i}k, N.~Loeff, and T.~Pfister, ``Temporal fusion transformers for interpretable multi-horizon time series forecasting,'' {\em International Journal of Forecasting}, vol.~37, no.~4, pp.~1748--1764, 2021.

\bibitem{2023_TSMixer}
S.-A. Chen, C.-L. Li, N.~Yoder, S.~O. Arik, and T.~Pfister, ``Tsmixer: An all-mlp architecture for time series forecasting,'' {\em arXiv preprint arXiv:2303.06053}, 2023.

\bibitem{open_data_elia}
{Elia}, ``{Open Data Elia}.''
\newblock \url{https://opendata.elia.be/pages/home/}.

\bibitem{dlinear}
A.~Zeng, M.~Chen, L.~Zhang, and Q.~Xu, ``Are transformers effective for time series forecasting?,'' in {\em 37th AAAI Conference on Artificial Intelligence}, vol.~37, pp.~11121--11128, 2023.

\bibitem{2015_ELU}
D.~Clevert, T.~Unterthiner, and S.~Hochreiter, ``Fast and accurate deep network learning by exponential linear units ({ELUs}),'' in {\em 4th International Conference on Learning Representations}, 2016.

\bibitem{2013_word2vec}
T.~Mikolov, K.~Chen, G.~Corrado, and J.~Dean, ``Efficient estimation of word representations in vector space,'' in {\em 1st International Conference on Learning Representations}, 2013.

\bibitem{2020_deep_NN_quantile_regression}
F.~Rodrigues and F.~C. Pereira, ``Beyond expectation: Deep joint mean and quantile regression for spatiotemporal problems,'' {\em IEEE Transactions on Neural Networks and Learning Systems}, vol.~31, no.~12, pp.~5377--5389, 2020.

\bibitem{adam}
D.~P. Kingma and J.~Ba, ``Adam: A method for stochastic optimization,'' {\em arXiv preprint arXiv:1412.6980}, 2014.

\bibitem{2019_ensemble_diversity_good}
T.~Pang, K.~Xu, C.~Du, N.~Chen, and J.~Zhu, ``Improving adversarial robustness via promoting ensemble diversity,'' in {\em International Conference on Machine Learning}, pp.~4970--4979, 2019.

\bibitem{2007_CRPS}
T.~Gneiting and A.~E. Raftery, ``Strictly proper scoring rules, prediction, and estimation,'' {\em Journal of the American statistical Association}, vol.~102, no.~477, pp.~359--378, 2007.

\end{thebibliography}

\end{document}